\documentstyle[psfig,epsf]{mn}

\title[Observations of pulsating subdwarf B stars]
{Resolving the pulsations of subdwarf B stars: PG~0154+182,
HS~1824+5745, and HS~2151+0857}

\author[M.D. Reed et al.]
{M. D. Reed,$^1$\thanks{E-mail:
MikeReed@missouristate.edu}
J. R. Eggen,$^1$ A.-Y. Zhou,$^1$ D. M. Terndrup,$^2$ S. L. Harms,$^1$  D. An,$^2$ 
\cr and  M. A. Hashier$^2$\\
$^1$Department of Physics, Astronomy and Materials Science, 
Missouri State University, 901 S. National, Springfield, MO 65897 USA \\
$^2$Department of Astronomy, The Ohio State University,
140 W. 18th Ave., Columbus, OH 43210 USA\\
}

\date{Accepted     
      Received }

\begin{document}

\maketitle

\begin{abstract}
We continue our programme of extended single-site observations 
of pulsting subdwarf B (sdB) stars and
present the results of extensive time series photometry 
to resolve the pulsation
spectra for use in asteroseismological analyses. 
PG~0154+182, HS~1824+5745, and HS2151+0857 were observed at
the MDM Observatory during 2004 and 2005. Our observations are sufficient
to resolve the pulsations of all three target stars.
We extend the number of known frequencies for PG~0154+182 from one to
six, confirm that HS~1824+5745 is a mono-periodic pulsator, and
extend the number of known frequencies to five for HS~2151+0857. 
We perform standard
tests to search for multiplet structure, measure amplitude variations
as pertains to stochastic excitation, and examine the mode density to
constrain the mode degree $\ell$.
\end{abstract}


\section{Introduction}

Subdwarf B (sdB) stars are thought to be stars with masses about 0.5~M$_{\odot}$
with thin ($<$$10^{-2}$M$_{\odot}$) hydrogen shells and temperatures
from $22\,000$ to $40\,000$~K 
(Heber et al. 1984; Saffer et al. 1994), making them exceedingly blue.
Pulsating sdB stars come in two varieties: short period (90 to 600
seconds) named EC~14026-2647 stars after that prototype, 
officially V361~Hya stars or
sdBV stars, with
amplitudes typically near 1\%; and long period (45 minutes to 2 hours) 
named PG~1716+426 stars after that prototype or LPsdBV stars, 
with amplitudes typically $<$0.1\%. For more
on these stars see Kilkenny (2001) and Green et al. (2003).
For this work, our interest is the EC~14026
(sdBV) class of pulsators as their periods are short, so many pulsation
cycles can be observed during each run from a single site. 

Pulsating sdB stars potentially allow the opportunity to discern
their interior structure using asteroseismology, obtaining estimates
of total mass, luminosity, shell mass,
radiative levitation, gravitational
diffusion, and helium fusion cross sections. However, to apply
the tools of asteroseismology, the pulsation frequencies (periods) must first 
be resolved.
Variable star discovery surveys seldom resolve or detect the complete set of
pulsations. Multisite campaigns, because of the complexity
of organization, have only observed a few sdB pulsators.

Our programme is to resolve poorly-studied sdB pulsators, typically
from single-site data. This method has proven useful
for the sdBV stars Feige~48 and KPD~2109+4401 (Reed et al. 2004; Zhou
et al. 2006). For these observations, we obtained most of our data at
a modest sized telescope (1.3~m) combined with higher signal-to-noise (S/N)
observations on a 2.4~m telescope. This combination allows us to
effectively resolve the pulsation spectrum of sdBV stars and detect
low amplitude pulsations that were below the detection limit of the
discovery data.
%and look for spacings which may provide mode identifications.
Our high S/N data from larger telescopes can detect
pulsation amplitudes as low as 0.5 milli-modulation amplitudes (mma, 
equivalent to 0.05\%), insuring that
we do not miss low amplitude modes. Our typical temporal resolution is
better than 1.5~$\mu$Hz. 

This paper reports the results of our follow-up observations on the sdBV stars
PG~0154+182 (hereafter PG~0154), HS~1824+5745 (hereafter HS~1824), and
HS~2151+0857 (hereafter HS~2151). PG~0154 ($B$=15.3) was reported as a
pulsator by Koen et al. (2004; hereafter K04) 
who detected a single frequency
at 6090~$\mu$Hz. Their observations consisted of three $<3$ hour 
observing runs which were obtained under less-than ideal conditions and
were certainly insufficient to resolve the pulsations. 
From optical spectroscopy and 2MASS infrared data, they also determined 
that PG~0154 has a main sequence G5 - K5 companion.
HS~1824 ($B$=15.6) and HS~2151 ($B$=16.5)  
were determined to be variable by \O stensen
et al. (2001; hereafter \O 01) as part of a four star discovery paper. 
(Observations of the
other two variables will be reported elsewhere.) They detected a single
frequency in HS~1824 and four in HS~2151. Their
observations consisted of three $\approx 1$ hour runs for HS~1824 and
two $\approx 1$ hour and one 2.5 hour run for HS~2151. As such, the
pulsations were likely unresolved and ideal for follow-up observations.
Both HS stars are apparently
single with $T_{\rm eff}=33~100~K$ and $34~500~K$ and
$\log g=6.0$ and $6.1$ for HS~1824 and HS~2151, respectively.

\section{Observations}
Data were obtained at MDM Observatory's
2.4~m and 1.3~m telescopes using an Apogee Alta U47+ CCD camera.
This camera is connected via USB2.0 for high-speed readout: Our
binned ($2\times 2$) images had an average dead-time of one second.
The observations used a red cut-off filter (BG38) so the effective
bandpass covers that of B and V filters and is essentially that
of a blue-sensitive photomultiplier tube. Such a setup allows us to
maximize light throughput while maintaining compatibility with
observations obtained with photomultiplier photometers. 
Tables~\ref{tab01} through \ref{tab03} provide the details of our
observations including date, start time, integration time and
run length.

PG~0154 was observed as a secondary target during our fall 2004 KPD~2109
data run (reported in Zhou et al. 2006). We obtained short
data runs ($<6.6$ hr) over nine nights.
HS~1824 was the subject of a concerted effort, with more than 125 hours of data
obtained during three different observing runs, two at the 1.3~m and
the last at the 2.4~m telescopes. After the first observing run on 
HS~1824, we began observing HS~2151 as a secondary target. More than 42 hours
of data were obtained of HS~2151 during three nights on the 1.3~m and
six on the 2.4~m telescopes. Though only the observations of HS~1824 conform
to our normal programme (long time-series observations covering several weeks),
we feel that we have successfully resolved all of the pulsations.

%Table 1
\begin{table}
\centering
\caption{Observations of PG~0154+182 during 2004 on the MDM 1.3~m 
telescope \label{tab01}}
\begin{tabular}{lcccc} \hline
Run & UT Start & Date & Length & Integration \\
 & hr:min:sec & UT & (hrs) & (s)\\ \hline
mdr277 & 02:50:30 & 06 Oct. & 3.7 & 20 \\
mdr279 & 02:50:30 & 07 Oct. & 3.8 & 20 \\
mdr281 & 02:50:30 & 08 Oct. & 3.7 & 20 \\
mdr283 & 02:50:30 & 09 Oct. & 4.2 & 20 \\
mdr286 & 02:50:30 & 10 Oct. & 1.2 & 20 \\
mdr288 & 02:50:30 & 11 Oct. & 4.1 & 20 \\
mdr291 & 02:50:30 & 12 Oct. & 0.4 & 20 \\
mdr293 & 02:50:30 & 13 Oct. & 6.6 & 20 \\
mdr296 & 02:50:30 & 14 Oct. & 0.8 & 20 \\ \hline
\end{tabular}
\end{table}

%Table 2
\begin{table}
\centering
\caption{Observations of HS~1824+5745 during 2005 \label{tab02}}
\begin{tabular}{lccccc} \hline
Run & UT Start & Date & Length & Int. & Telescope \\
 & hr:min:sec & UT & (hrs) & (s) & \\ \hline
mdm052505 & 04:35:00 & 25 May & 6.9 & 20 & 1.3~m \\
mdm052605 & 04:12:00 & 26 May & 7.4 & 20 & 1.3~m \\
mdm052705 & 08:32:00 & 27 May & 3.1 & 20 & 1.3~m \\
mdm053005 & 04:13:30 & 30 May & 7.4 & 20 & 1.3~m \\
mdm053105 & 04:43:00 & 31 May & 6.9 & 20 & 1.3~m \\
mdm060105 & 04:13:30 & 01 June & 7.4 & 20 & 1.3~m \\
mdm060305 & 03:37:00 & 03 June & 7.9 & 20 & 1.3~m \\
mdm060405 & 03:48:30 & 04 June & 7.7 & 20 & 1.3~m \\
mdm060505 & 03:48:30 & 05 June & 7.8 & 20 & 1.3~m \\
mdm060605 & 03:45:00 & 06 June & 7.8 & 20 & 1.3~m \\
mdm061305 & 04:17:11 & 14 June & 6.6 & 20 & 1.3~m \\
mdm061505 & 03:50:00 & 15 June & 7.6 & 20 & 1.3~m \\
mdm061605 & 03:33:00 & 16 June & 8.1 & 20 & 1.3~m \\
hs18\_061705 & 03:36:00 & 17 June & 3.0 & 20 & 1.3~m \\
hs18\_061905 & 03:33:00 & 19 June & 1.9 & 30 & 1.3~m \\
hs18\_062005 & 03:26:00 & 20 June & 4.4 & 25 & 1.3~m \\
hs18\_062205 & 05:40:00 & 22 June & 5.9 & 25 & 1.3~m \\
hs18\_070605 & 04:17:00 & 06 July & 1.2 &  7 & 2.4~m \\
hs18\_070705 & 04:14:00 & 07 July & 1.5 &  7 & 2.4~m \\
hs18\_070805 & 03:21:00 & 08 July & 5.4 &  7 & 2.4~m \\
hs18\_070905 & 03:18:00 & 09 July & 2.2 &  7 & 2.4~m \\
hs18\_071005 & 03 15 30 & 10 July & 2.1 &  7 & 2.4~m \\
hs18\_071105 & 03:20:50 & 11 July & 4.1 &  7 & 2.4~m \\
hs18\_071205 & 03:23:30 & 12 July & 3.1 &  7 & 2.4~m \\ \hline
\end{tabular}
\end{table}

%Table 3
\begin{table}
\centering
\caption{Observations of HS~2151+0857 during 2005 \label{tab03}}
\begin{tabular}{lccccc} \hline
Run & UT Start & Date & Length & Int. & Telescope \\
 & hr:min:sec & UT & (hrs) & (s) &\\ \hline
hs21\_061805 & 09:10:00 & 18 June & 2.1 & 25 & 1.3~m \\
hs21\_061905 & 07:40:00 & 19 June & 3.9 & 26 & 1.3~m \\
hs21\_062005 & 08:06:00 & 20 June & 3.5 & 28 & 1.3~m \\
hs21\_070605 & 06 01 00 & 06 July & 5.8 & 10 & 2.4~m \\
hs21\_070705 & 05:55:50 & 07 July & 5.9 & 10 & 2.4~m \\
hs21\_070805 & 08:55:00 & 08 July & 2.9 & 10 & 2.4~m \\
hs21\_070905 & 05:43:40 & 09 July & 6.1 & 10 & 2.4~m \\
hs21\_071005 & 05:27:00 & 10 July & 6.4 & 10 & 2.4~m \\
hs21\_071205 & 06:33:41 & 12 July & 5.2 & 10 & 2.4~m \\ \hline
\end{tabular}
\end{table}

\begin{figure*}
\centerline{\psfig{figure=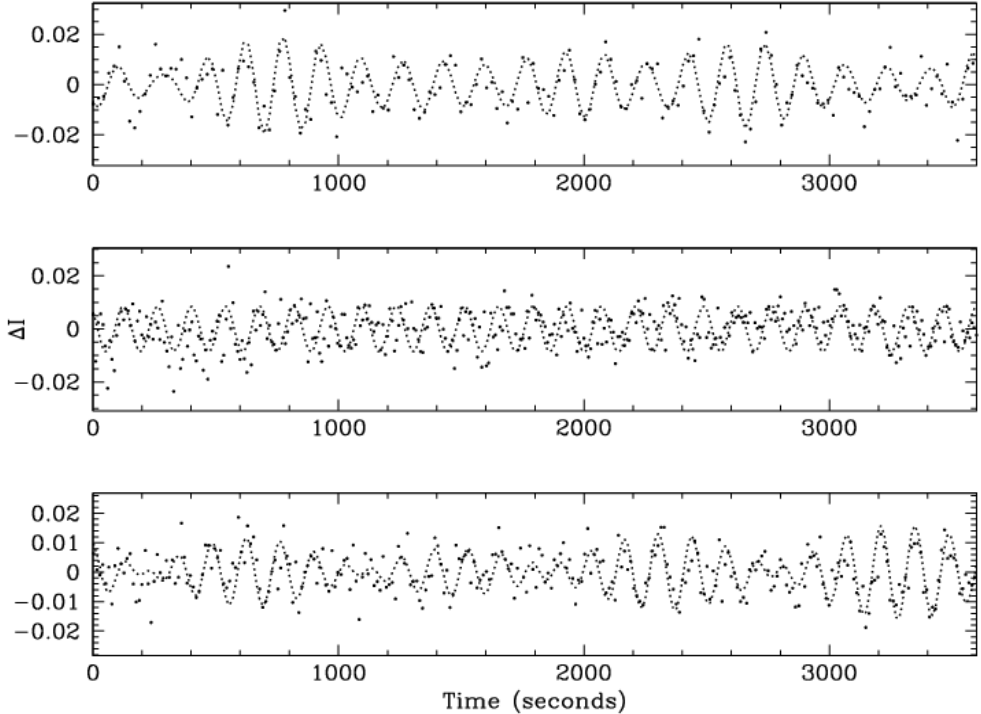,width=5.5in}}
\caption{Sample one-hour lightcurves of the three targets. From top to
bottom are PG~0154 (obtained on the 1.3~m), HS~1824 (obtained on the 2.4~m), 
and HS~2151 (obtained on the 2.4~m). The dotted line is our least-squares
solution of the entire data sets.}
\label{fig01}
\end{figure*}

Standard procedures of image reduction, including bias subtraction,
dark current and flat field correction, were followed using 
IRAF\footnote{IRAF is distributed by the National Optical Astronomy 
Observatories, which are operated by the Association of Universities 
for Research in Astronomy, Inc., under cooperative agreement with the 
National Science Foundation.} packages.
Intensities were extracted using IRAF aperture photometry with extinction
and cloud corrections using the normalized intensities of several field
stars. As sdB stars are substantially hotter than typical field
stars, differential light curves
are not flat due to differential atmospheric and colour extinctions.
A low-order polynomial was fit to remove these trends from the data on a
night-by-night basis. Finally, the lightcurves are normalized by their
average flux and centred around zero so the reported differential
intensities are $\Delta {\rm I}=\left( {\rm I}/\langle{\rm I}\rangle\right)-1$.
Amplitudes are thus given as milli-modulation amplitudes (mma) with an
amplitude of 10~mma corresponding to 1.0\% or 9.2~millimagnitudes.
Sample lightcurves are shown in Figure~\ref{fig01}.

\section{Pulsation Analysis}
\begin{figure*}
\centerline{\psfig{figure=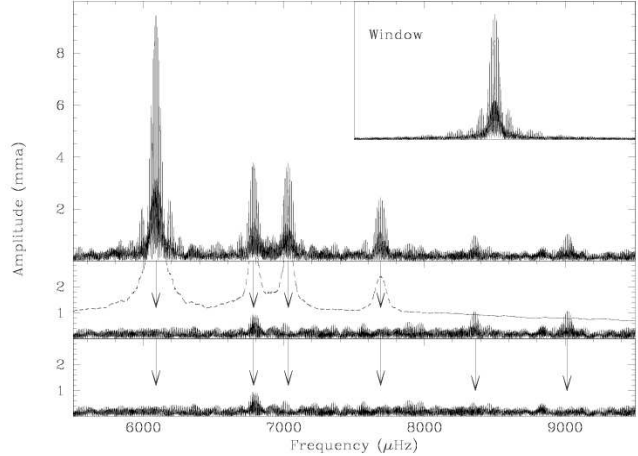,width=3.5in}}
\caption{Temporal spectra (FT) for PG~0154 showing the original data (top),
and prewhitening by 4 and 6 frequencies (middle and bottom panels,
respectively). Arrows indicate
prewhitened frequencies while the dashed line in the middle panel shows 
the $4\sigma$ detection limit. Inset is the data window plotted at the
same horizontal scale.}
\label{fig02}
\end{figure*}

{\bf PG~0154: }
The temporal spectrum and window function
 for PG~0154 are plotted in Figure~\ref{fig02}. The top panel shows a Fourier
transform (FT; also known as temporal or pulsation spectrum) of the 
original data, with the two lower panels showing the
effects of prewhitening by four and six frequencies, respectively. 
The inset is the window function, which is a
single, noise-free sine wave (of arbitrary amplitude) 
sampled at the same times as the data. The central peak of the window 
is the input frequency with other peaks indicating the
aliasing pattern of the data.
In order to determine the significance of the two low-amplitude peaks,
we convolved a slightly smoothed window function which included the four
highest amplitude peaks with the data noise as determined using the
Breger et al. (1994; hereafter B94) criterion, which is $4\times$ 
the average
value of the FT (the dashed line in Figure~\ref{fig02}). Using such a 
conservative estimate of the noise, we 
can say with confidence that both low-amplitude
peaks are real and significantly above the noise. Frequencies,
amplitudes and phases were determined by simultaneously fitting a non-linear
least-squares solution to the data. Our solution for the frequencies
and amplitudes
is provided in Table~\ref{tab04}. No other frequencies were detected
up to the Nyquist frequency near 25000~$\mu$Hz. 

We readily confirm the frequency detected
by K04 and detect five additional pulsation frequencies.
 As the closest two frequencies are separated
by nearly 300~$\mu$Hz, the pulsations in PG~0154 can readily be resolved in
about 1 hour of observations, a very fortunate circumstance since our
data runs were short in duration. We also examined 
individual observing runs of sufficient length and readily detected all
six pulsation frequencies. Though there was some amplitude variability
(to be discussed in \S 4.3) which is responsible for the peak in
the residuals at $6784\,\mu$Hz, the frequencies were stable over the course
of our observing runs.

\begin{table}
\caption{Periods, frequencies, and amplitudes for PG~0154. Formal
least-squares errors are in parentheses. }
\label{tab04}
\begin{center}
\begin{tabular}{lccc}
\hline
ID & Period & Frequency  & Amplitude \\
 & (s) & ($\mu$Hz) & (mma) \\
$f1$ & 110.9305 (18)  &  9014.65 (15) & 1.1 (2) \\
$f2$ & 119.5827 (22)  &  8362.09 (15) & 1.1 (2) \\
$f3$ & 130.0694 (10)  &  7688.20 (06) & 2.5 (2) \\
$f4$ & 142.2167 (08)  &  7031.52 (04) & 3.9 (2) \\
$f5$ & 147.3860 (09)   &  6784.90 (04) & 3.9 (2) \\
$f6$ & 164.2108 (03)  &  6089.73 (01) & 9.5 (2) \\ \hline
\end{tabular}
\end{center}
\end{table}

\begin{figure*}
\centerline{\psfig{figure=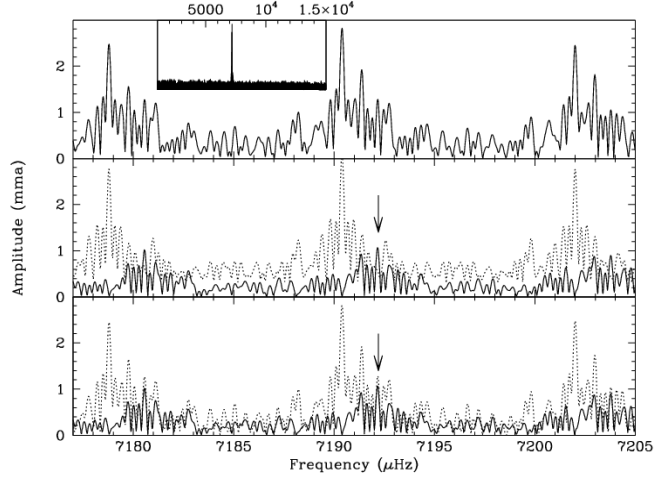,width=3.5in}}
\caption{Temporal spectra (FT) for HS~1824 showing the pitfalls of
prewhitening. The top panel shows the original FT (the inset spans a 
larger region) and the middle panel shows the original FT prewhitened by 
one frequency (solid line) using a window function fit from the entire
data set (dashed line). The bottom panel shows the same FT as the middle
panel, but the window function incorporates the amplitude and phase 
variations determined from individual nights of data.}
\label{fig03}
\end{figure*}

{\bf HS~1824:}
The majority of our 2005 campaign was devoted to confirming that
HS~1824 has just a single pulsation frequency. The formal least squares
solution for the frequency, using all of the data is 
$f=7190.413397\pm  0.000005\,\mu$Hz which is on the low end of 
\O 01, but within their errors. Using the B94 criterion provides
a limit of undetected pulsations at 0.48~mma outside of the window pattern
caused by the single peak. Additional peaks were searched for
up to the Nyquist frequency near 50000~$\mu$Hz, but none were
detected. However, just because
HS~1824 is mono-periodic does not make it uninteresting. It has both 
amplitude and phase variations that will be discussed in \S 4. The average
amplitude of the pulsation was 2.4~mma over the course of our observations,
but ranged from 2 to 6~mma. Such amplitude variation points out a weakness
in the prewhitening process. Prewhitening algorithms typically remove 
a constant amplitude
sinusoidal function from the data at the frequency, amplitude,
and phase determined via non-linear least-squares fitting (or several
for multiple pulsations). Figure~\ref{fig03} shows how this can affect
the prewhitening process. A 30~$\mu$Hz region centred on the pulsation
frequency is shown for HS~1824. The top panel shows the original FT (the
inset shows a 14,000~$\mu$Hz span) and the middle panel shows the FT 
prewhitened by one frequency (solid line) over-plotted on the window 
function (dashed line) that matches the entire data set. The arrow 
points to a peak that is still above the window function, which could
be (mistakenly) deemed intrinsic to the star. The bottom panel shows the
same prewhitened FT as the middle panel, except that the window
function is created using the least-squares information from individual
runs. In this case, there are no peaks left above the window and we
correctly deduce the mono-periodic nature of the pulsations.

\begin{figure*}
\centerline{\psfig{figure=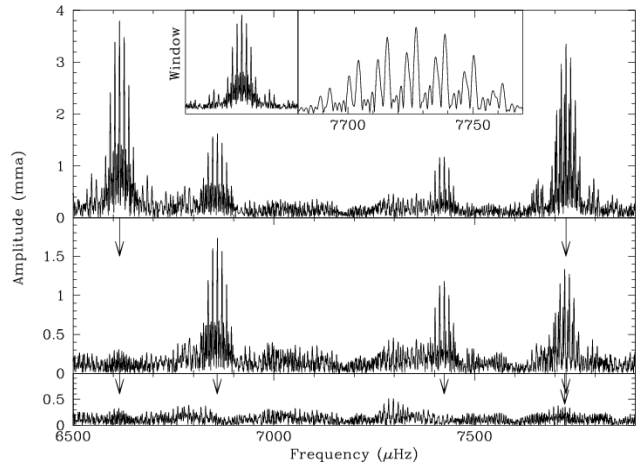,width=3.5in}}
\caption{Temporal spectra (FT) for HS~2151 showing the original data (top),
and prewhitening by 2 (middle) and 5 frequencies
(bottom). Arrows indicate
prewhitened frequencies. Inset shows the data window and an
expanded view of the
close pair of frequencies.}
\label{fig04}
\end{figure*}

{\bf HS~2151:}
The majority of our data for HS~2151 was obtained during our 
observing run in July, 2005, with just 
three relatively short runs obtained during
the late June run. As such, we examined these data as three separate
sets: The June data, the July data, and the combined data. The June data alone
are not sufficient to resolve the close doublet near 7725~$\mu$Hz, and
the combined data have a slightly more complex window function then just
the July data. The B94 noise criterion for the combined and July runs are
within 0.02~mma of each other (0.51 and 0.53~mma, respectively), 
and so the best results were obtained using the July data only.
The temporal spectrum of HS~2151 is shown in Figure~\ref{fig04}. The
top panel shows the original FT of the July data 
with the next two panels showing the
prewhitening sequence of two and five frequencies, respectively. The
insets show the window function and an enlarged section of the original
FT to show the resolved doublet. Arrows indicate the prewhitened 
frequencies and our least-squares solution is provided in Table~\ref{tab05}.
No further frequencies were detected up to the Nyquist frequency near
39000~$\mu$Hz.

\begin{table}
\caption{Periods, frequencies, and amplitudes for HS~2151. Formal
least-squares errors are in parentheses. }
\label{tab05}
\begin{center}
\begin{tabular}{lccc}
\hline
ID & Period & Frequency  & Amplitude \\
& (s) & ($\mu$Hz) & (mma) \\
$f1$ & 129.4133 (07) &   7727.18 (4) & 3.01 (13) \\
$f2$ & 129.4671 (13) &   7723.97 (8) & 1.51 (13) \\
$f3$ & 134.6929 (17) &   7424.30 (9) & 1.20 (12) \\
$f4$ & 145.7850 (13) &   6859.41 (6) & 1.74 (12) \\
$f5$ & 151.1561 (07)  &  6615.68 (3) & 3.84 (12) \\
\\ \hline
\end{tabular}
\end{center}
\end{table}

\section{Discussion}
\subsection{Comparison with the discovery data}
The goal of these observations was to obtain better data than in the
discovery papers and to resolve the pulsation frequencies for
asteroseismic analysis. For the sake of comparison with the discovery data,
we calculate the temporal resolution as $1/\Delta T$ with $\Delta T$ being
the temporal extent of the observations. For the discovery data, we can 
determine temporal resolution from
information provided in their papers, and 
estimate ``by eye'' the discovery data detection limit as twice the top of 
the noise level in their FTs outside of the pulsations and windows. 

For PG~0154, we were able to confirm the frequency
from K04 as well as detect five new frequencies. Our observations have a
temporal resolution of 1.4~$\mu$Hz which is more than $4\times$ better than
the discovery data and a detection limit of 0.76~mma which is 
more than $6\times$
better than in K04.

We certainly obtained more and better data than \O 01 for HS~1824, but it
ended up only confirming the singe frequency they detected. Our temporal
resolution is 0.25~$\mu$Hz, which is $23\times$ better than theirs and
our detection limit is 0.48~mma which is $\approx 4\times$ better than
\O 01.

For HS~2151 we confirm the four previously 
detected frequencies of \O 01 and detect
an additional, closely spaced frequency. This detection was possible 
because of our $11\times$ better temporal resolution of 0.5~$\mu$Hz and our
detection limit was also $\approx 3\times$ better than \O 01 at 
0.53~$\mu$Hz.

Compared to the discovery data of these three stars, we certainly obtained
our goal of vastly superior temporal resolution and 
improved detection limits. However, it is interesting to note that 
both PG~0154 and HS~1824 showed a single peak in very limited  
discovery data, and while PG~0154 paid dividends with only relatively
few observations, no new pulsations were detected in HS~1824 after a 
concerted effort.

\subsection{Multiplet constraints on pulsation modes}
For our goal of applying asteroseismological tools to these stars, one of
the most helpful features would be observational constraints on the
pulsation \emph{modes} themselves, rather than just recording the
pulsation frequencies. Pulsation modes can be described by three quantum
numbers, $n$ (or $k$), $\ell$, 
and $m$, and mathematically described by spherical
harmonics. As rotation can break the $m$ degeneracy by separating each
degree $\ell$ into a multiplet of $2\ell +1$ components, multiplet
structure is a very useful tool for observationally constraining pulsation
degree (see Winget et al. 1991). Such multiplets should be nearly
equally spaced in frequency but such structure is seldom observed in
sdBV stars. However, when multiplets are observed, they not only
constrain the pulsation degree ($\ell$) but also the rotation period
and inclination (Reed et al. 2004; O'Toole, Heber, \& Benjamin 2004).

For PG~0154
there is a common value that emerges with an average
frequency spacing of $\approx 670\mu$Hz that includes \emph{all} of the
observed frequencies. The spacings are provided in 
Table~\ref{tab06} and range from 652 to
695~$\mu$Hz. This is less than a 5\% variation and roughly of the 
order expected from theory. \emph{If} these are rotationally split
multiplets, $f1$ through $f4$ would necessarily belong to a single degree
of $\ell \geq 2$, missing at least one component (just one for $\ell =2$)
and $f5$ and $f6$ would be two parts of an $\ell \geq 1$ multiplet. As
such, all of the frequencies in PG~0154 can be explained using one
$\ell =1$ and one $\ell =2$ mode; each missing one of their components.
However, rotational splitting of 670~$\mu$Hz means a rotation period
of 1492~s, or about 25 minutes! At a canonical radius of 
$0.15{\rm R}_{\odot}$, this would imply a rotation velocity of 
$\approx 400$ km/s; an easily measurable quantity. Spectroscopic
constraints already exist and from Figure~7 of K04, it is obvious that 
PG~0154 is a normal, slowly rotating sdB star. Though it could be
possible to conceal such a high velocity with a low inclination, another
argument against this scenario is the orbital separation. Using the
canonical sdB mass of 0.5~M$_{\odot}$ and a range of masses appropriate
for the main sequence companion, the orbital distance between the two
mass centres is $<$0.95~R$_{\odot}$. The two stars could not exist
within such close proximity to each other without significant effects;
at a minimum, the sdB star would have significant ellipsiodal variations 
like those observed for the close binary KPD~1930+2752 (Bill\` eres et  
al. 2000) and for most companion parameters, the two stars would not
fit within the orbital separation.

Since the spacing is so large, it is conceivable that they could be 
successive overtones of the radial index, $n$. However, because $f4$ and
$f5$ do not share this spacing, they would necessarily belong to differing
degrees, $\ell$, and thoery predicts that  differing degrees
have different
overtone spacing. However, a glance at Figure~4 of
Charpinet et al. (2002) reminds us that theory expects these to be low
overtone pulsations, which do not obey the asymptotic relationship of 
even frequency spacing. The figure indicates that successive overtones
should \emph{differ} in spacing by at least 500~$\mu$Hz, ruling out this
possibility.

% Table 6
\begin{table}
\centering
\caption{Frequency spacings between pulsation frequencies for PG~0154.
\label{tab06}}
\begin{tabular}{cc}
Frequency pair & Difference \\
 & ($\mu$Hz)\\ \hline
$f1-f2$ & 652.6  \\
$f2-f3$ & 673.9  \\
$f3-f4$ & 656.7  \\
$f5-f6$ & 695.4  \\ \hline
\end{tabular}
\end{table}

\begin{figure*}
\centerline{\psfig{figure=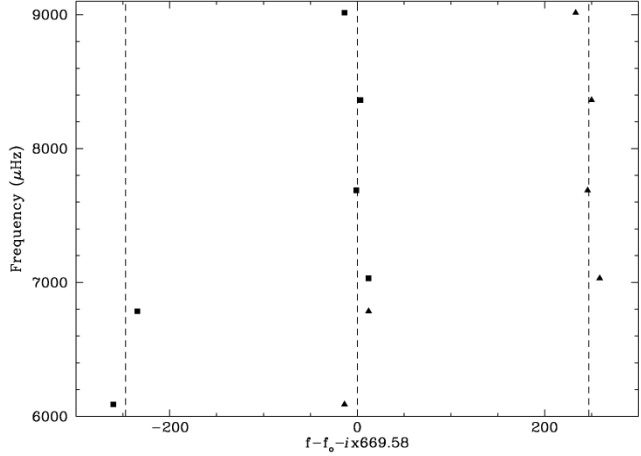,width=3.5in}}
\caption{Echelle diagram for PG~0154 using ``Kawaler scheme'' model
fits. Squares correspond to the model with $f_o=6349.9\,\mu$Hz and triangles
for the model with $f_o=6103.2\,\mu$Hz.}
\label{fig04b}
\end{figure*}

Another possibility is the ``Kawaler scheme'' which has recently been
introduced (Kawaler et al. 2006; Vu\u{c}kovi\'{c} et al. in preparation).
Though it is
not compatible with low overtone ($n$) pulsation theory invoked for sdBV
stars, it has been noticed that for several stars, an improved
frequency fit can be obtained using an asymptotic-like formula;
$$ f(i,j)=f_o+i\times\delta +j\times\Delta$$
where $i$ has integer values, $j$ is limited to values of $-1,\,0,$ and $1$,
and up till now $\delta$ has been a small spacing and $\Delta$ has
been a large spacing. However, for PG~0154, the large spacing occurs too
frequently to be restricted to just three values, ($-1,\,0,\,+1$). 
For PG~0154, there are not enough frequencies to make a unique fit and
so Table~\ref{tab07} gives the solutions for two possible values of
$f_o$ and Figure~\ref{fig04b} shows the Echelle diagram of the fits
using a different symbol for each fit. (Fractional multiples of the
$\delta$ spacings, i.e. $\delta /2$, are also equally good fits but the
largest spacing comes directly from the data.) 
For both models, $\delta =669.6\mu$Hz
and $\Delta =247.3\mu$Hz using $f_o=6349.9\mu$Hz for the squares and
$f_o=6103.2\mu$Hz for the triangles. Part of the Kawaler scheme is to
choose  ``the ($\delta ,\,\Delta$) pair for which we have at least
two modes with the same value of $i$ but different $j$'' and we have
satisfied this condition, but without coming to a unique conclusion.
Still, we can say that the Kawaler scheme fits the data with an r.m.s. error
of only 0.16\% and remains an option while a rotational explanation for the
frequency splitting is clearly ruled out.
A Kawaler scheme fit was
also obtained switching the values for $\delta$ and $\Delta$, but the
differences were substantially larger.

% Table 6
\begin{table}
\centering
\caption{A ``Kawaler scheme'' model fit to the observed frequencies using
$f_o=6349.9$ (6103.2), $\delta =669.6$ and $\Delta =247.3\mu$Hz.
\label{tab07}}
\begin{tabular}{lcccc} \hline
Star & $i$ & $j$ & Model & Difference  \\ \hline
$f1$ & 4 & 0 (1) & 9028.1 & -13.5 \\
$f2$ & 3 & 0 (1) & 8358.6 & 3.5 \\
$f3$ & 2 & 0 (1) & 7689.0 & -0.8 \\
$f4$ & 1 & 0 (1) & 7019.4 & 12.1 \\
$f5$ & 1 & -1 (0) & 6772.5 & 12.4 \\
$f6$ & 0 & -1 (0) & 6103.4 & -13.1 \\ \hline
\end{tabular}
\end{table}

For HS~2151, where we have five frequencies to work with, no two 
frequency spacings are similar. As such, we are left with the following
possibilities for the five pulsation frequencies detected: 1) Rotation
is sufficiently slow that all $m$ values are degenerate; 2) at most one
pair share the same $n$ and $\ell$ values with differing $m$, which seems
very unlikely; 3) our line of sight is along the pulsation axis,
with $\sin i\approx 0$, making only $m=0$
modes observable because of geometric cancellation (Pesnell 1985; Reed,
Brondel, \& Kawaler 2005); or 4)
internal rotation is such that rotationally-induced multiplets are 
widely spaced and uneven (Kawaler \& Hostler 2005).

\subsection{Constraints on pulsation degree via mode density}
An open question involving sdBV stars is the mode degree $\ell$ of the
pulsations. In resolved sdBV stars, we sometimes observe many more 
pulsation modes than $\ell$=0, 1, and 2 can provide. Higher $\ell$ modes
may be needed, but if so they must have a larger intrinsic
amplitude because of the large degree of geometric cancellation 
(Charpinet et al. 2005; Reed, Brondel, \& Kawaler 2005). A general
guideline would be one $n$ order per $\ell$ degree per 1000~$\mu$Hz. As
such, the temporal spectrum can accommodate three frequencies
per 1000~$\mu$Hz without the necessity of invoking high-$\ell$ values.

For PG~0154, even ignoring the frequency spacing that inter-relates several
frequencies, it is still easy to accommodate all the frequencies without
the need for $\ell\geq 3$ modes. The pulsation spectrum for HS~2151 is
slightly more dense and we do not detect any related frequencies. However,
it is still possible to fit the frequencies within the range observed without
invoking $\ell\geq 3$ modes if the 6616 and 6859~$\mu$Hz frequencies are
related to those near 7725~$\mu$Hz. As such, we can deduce that the pulsations
in all three pulsators in this paper \emph{can} be accommodated using
$\ell =0$, $1$, and $2$ modes. However, this by no means implies that they
are.

\subsection{Amplitude variability}
Another feature we can examine is the amount of amplitude variability in 
the pulsations. If pulsating sdB 
stars are observed over an extended time period,
it is common to detect amplitude variability in many, if not all, of the
pulsation frequencies.
Such variability can occasionally be ascribed to beating between 
pulsations too
closely spaced to be resolved in any subsets of the data, but often
appear in clearly resolved pulsation spectra where
mode beating cannot be the cause. 

\begin{figure*}
\centerline{\psfig{figure=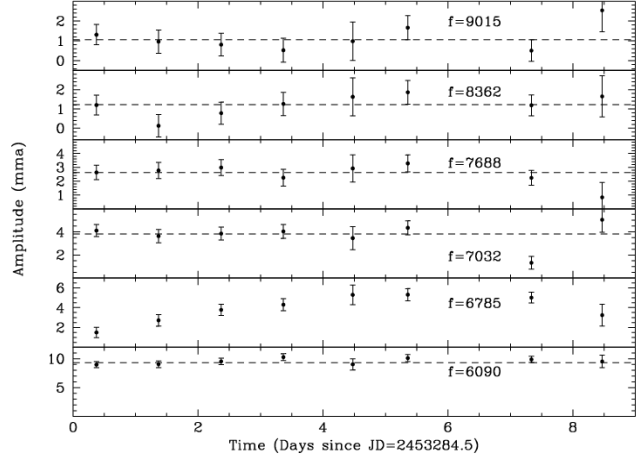,width=3.5in}}
\caption{Pulsation amplitudes for frequencies detected in PG~0154.
Error bars are the formal least--squares errors.}
\label{fig05}
\end{figure*}

Figure~\ref{fig05} shows the amplitudes of the pulsations of PG~0154 for
individual nights. The scale is selected (for all of the amplitude and
phase plots) such that a 10\% errorbar occupies $\approx$10\% of the
panel.
All of the amplitudes are essentially constant to within the errors 
except for $f5$,
which triples in amplitude over the course of the run. We initially
suspected it was caused by an unresolved doublet of frequencies, which
could still be the case. However, several other frequencies show the
same general trend (though on smaller scales) and the amplitude
variations do not appear sinusoidal, which would be expected with mode
beating. As such, we doubt our initial
interpretation and rather think these variations are intrinsic. 
However, the only way to be sure is to obtain more
observations. Still, the net result is that five of six frequencies are
stable both in amplitude and phase (not shown as all of the phases
are constant to within the errors).

\begin{figure*}
\centerline{\psfig{figure=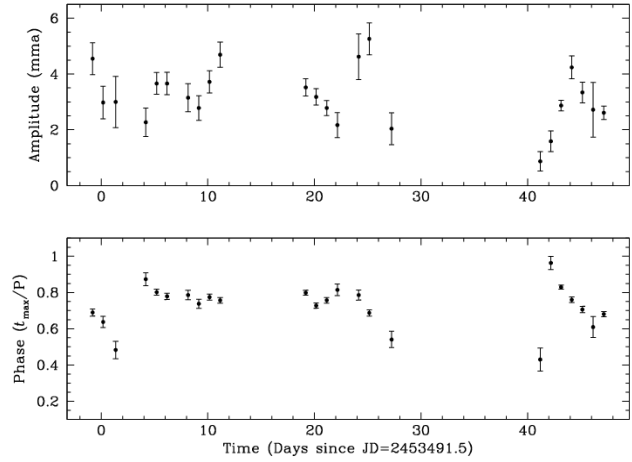,width=3.5in}}
\caption{Pulsation amplitudes and phases for frequencies detected in HS~1824.
Error bars are the formal least--squares errors.}
\label{fig06}
\end{figure*}

Of significantly more interest are the amplitudes and phases of the 
single frequency present in HS~1824. These are shown in Figure~\ref{fig06}
and indicate that they are quite variable. Between observing days 41 and
45 (JD=2453531.5 and 2453536.5), the amplitude rises from 0.9 to 4.2~mma
while the phases change by $\approx 40$\% between days 42 and 46.
Clearly, HS~1824 is both amplitude and phase variable; 
possibly an indicator
of stochastic pulsations (discussed in the next section).
Note that phases  are chosen such that the
phase is the time of first maximum since JD=2453279.5 for PG~0154,
JD=2453491.5 for HS~1824 and HS~2151 divided by the period.

\begin{figure*}
\centerline{\psfig{figure=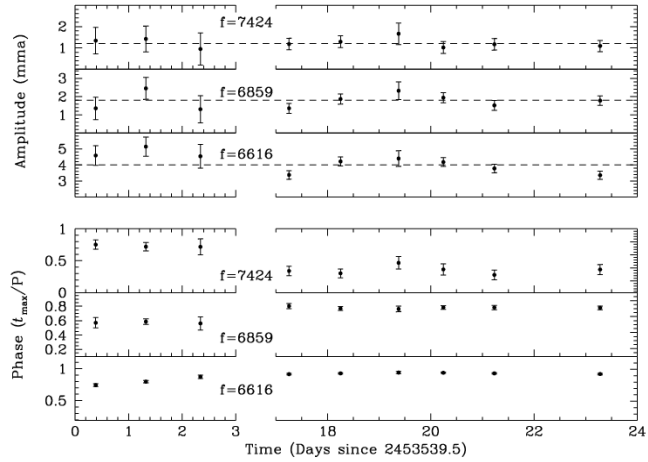,width=3.5in}}
\caption{Pulsation amplitudes and phases for frequencies detected in HS~2151.
Error bars are the formal least--squares errors.}
\label{fig07}
\end{figure*}

Figure~\ref{fig07} shows the amplitudes and phases of the three well
separated frequencies of HS~2151. The doublet near 7725~$\mu$Hz is not
resolvable on a nightly basis. In this case, two of the three frequencies
($f4$ and $f5$) have amplitude variations marginally
larger than the errors which appear to 
follow the same trend. One possible explanation would be if they
are two parts of a multiplet sharing an increase in driving power, 
though such a hypothesis cannot be tested  using
these data. For all three frequencies, the phases vary over
the duration of our observing campaign. The highest amplitude frequency
($f5$) is generally the most stable; varying smoothly by 15\% over
the first three nights and then remaining essentially constant. The other
two frequencies ($f3$ and $f4$) both show an offset between the June and 
July data runs with $f4$ essentially constant during the individual
month's data. This points toward a problem with timing rather than intrinsic
changes of the star. We investigated this possibility but ruled it out for
the following reasons: The highest amplitude frequency does not show these,
the changes are not the same in period (i.e., the same amount of difference
in timing), our observations of HS~1824, which occurred during the same
nights do not show this shift (though admittedly HS~1824 shows phase
variations), and the timing setup is the same as we have used during all of
our other runs\footnote{The acquisition computer's clock is synchronized
with time servers through Network Time Protocol.} 
and is checked multiple times during the night. 
As such, we conclude that these shifts that occurred during the 15 night
interval between observations are intrinsic to the star.

\subsection{The nature of the excitation mechanism}
We can also use the criterion outlined in Christensen-Dalsgaard et al (2004;
hereafter JCD04)
and first applied to sdBV stars in Pereira \& Lopes (2005) to 
examine the nature of the excitation mechanism.
If the frequencies are stochastically excited, we can expect to find
the standard deviation of the amplitudes, $\sigma (A)$, divided by the
average amplitude $\langle A\rangle$ to 
have a value near 0.5 (Eqn. 7 of Pereira \& Lopes 2005).

Both parameters and their ratios are given in Table~\ref{tab08} for all
resolved frequencies in our target stars.
For PG~0154, three of the six frequencies have relatively stable
pulsation amplitudes, and their ratios reflect this as non-stochastically
excited pulsations. The two low amplitude frequencies, $f1$ and $f2$, 
have errors too large to constrain the ratio, which for
$f5$ is in between. In this case $f5$ suffers the same problem as it did
for mode beating: The time-scale of amplitude variation appears longer
than our observations. As such, more data would be required to discern
which interpretation is correct for $f5$: Stochastically driven pulsations,
mode beating, or variable amplitudes caused by something else.
Surprisingly the ratio for HS~1824
is nearly the same as for $f5$ of PG~0154.  However
we can suggest that
these pulsations do not match the JCD04 criteria for stochastic oscillations 
as our observations are
surely sufficient in duration to include several damping lifetimes, yet
$\sigma (A)/\langle A\rangle$ is only $0.31\pm 0.11$. As such, the
rather large variation in amplitude and phase remains unexplained.
For the three well-separated frequencies of HS~2151,
none of them have ratios indicating stochastically excited pulsations. These
results indicate that at most one of our ten resolved frequencies have
indications of being stochastically excited. This can be compared to 
our previous observations of KPD~2109+4401 where 50\% of the resolved
pulsations indicated stochastic excitation and a study of PG~1605+072
where \emph{no} indications of stochastic excitation were found from
11 resolvable frequencies (Pereira \& Lopes 2005). It is a bit early
to interpret the meaning of such results and it is not clear what impact
the discovery of many stochastically driven frequencies (should that
occur) would mean for
the proposed iron driving mechanism of Charpinet et al. (2001).

% Table 8
\begin{table}
\centering
\caption{Mean amplitude, standard deviation, and their ratio for 
readily resolvable frequencies. \label{tab08}}
\begin{tabular}{cccc}
Frequency & $\langle A\rangle$ & $\sigma (A)$ &$\sigma (A)/\langle A\rangle$ \\
 ($\mu$Hz) & (mma) & (mma) & \\ \hline
\multicolumn{4}{c}{PG~0154}\\
9015 & 0.97  & 0.38 & $0.39\pm 0.39$ \\
8362 & 1.37  & 0.34 & $0.25\pm 0.22$ \\
7688 & 2.72  & 0.36 & $0.13\pm 0.12$ \\
7032 & 4.07  & 0.49 & $0.12\pm 0.05$ \\
6785 & 3.90  & 1.27 & $0.33\pm 0.06$ \\
6090 & 9.53  & 0.48 & $0.05\pm 0.03$ \\ \hline
\multicolumn{4}{c}{HS~1824}\\
7190 & 3.19 & 1.01 & $0.31\pm 0.11$ \\ \hline
\multicolumn{4}{c}{HS~2151}\\
7424 & 1.24 & 0.16 & $0.17\pm 0.12$ \\
6859 & 1.77 & 0.40 & $0.22\pm 0.10$ \\
6616 & 4.18 & 0.55 & $0.13\pm 0.04$ \\ \hline
\end{tabular}
\end{table}

\section{Conclusions}
From extensive follow-up data acquired at MDM observatory, we are
confident that we have resolved the pulsation spectra of three additional
pulsating sdB stars. We are able to confirm all of the previously observed
frequencies as well as detect a total of six new frequencies in two of
our three targets. 

For PG~0154, we have discovered five new 
frequencies with the closest spacing between frequencies at
$\approx 250\,\mu$Hz. As such, they are readily resolvable in 1.5 hours
of data, however the low amplitude of two frequencies requires high S/N
or longer time-base observations to reduce the noise. We notice an enticing
common frequency spacing of $\approx 670\,\mu$Hz, which could be 
interpreted as all six frequencies being members of one $\ell =1$ and
one $\ell =2$ pulsation multiplet. However, this would require a rotation
period near 25 minutes and a rotation velocity of $\approx 440$~km/s
which is clearly ruled out by K04. We applied the Kawaler scheme to
the spacings using
a purely numerical fit with good results. However, as PG~0154 only
has six frequencies, there is no unique solution and so we are not convinced
that it should be applied in this case. As such, the
cause for the large, nearly-equal frequency spacing remains a mystery.
Three of the six frequencies are 
amplitude-stable, while one shows variations that could be indicative of
a stochastically excited mode. Of course this would not be possible if
it is actually an $m$ component of the same degree as other,
amplitude-stable frequencies. As such, it seems we still have some to
learn about PG~0154 and additional follow-up observations could still
be instructive.

In HS~1824 we confirmed the frequency detected by \O 01, but did not
uncover any additional frequencies. However, we were able to measure
significant amplitude and phase variations over a long span of time.
As such, once asteroseismology matures to where it can
interpret such fluctuations, HS~1824 will be an excellent target. But
until then, all we can readily say is that the
variability in the amplitude and phase of pulsation seems unrelated
to stochastically excited oscillations.

For HS~2151, we confirm the four previously known frequencies and detect
an additional oscillation close to a previously observed one. However,
none of the frequencies appear to have common spacing indicative of 
multiplet structure and their amplitudes appear stable,
indicating non-stochastically excited pulsations.

These three pulsating sdB stars show the variety, and perhaps the complexity,
of oscillations observed in sdB stars. Two of the three have several 
frequencies, while HS~1824 only had one. Of the ten total frequencies,
several show amplitude variations, but only one has sufficient variability
to even consider stochastic excitation. Several of the phases show small
variations with those in
HS~1824 being much larger and only one star shows a common frequency spacing
that could be used to observationally constrain the pulsation degree,
$\ell$, except it would require a drastically fast rotation velocity 
which has been ruled out by previous spectroscopic observations.

ACKNOWLEDGMENTS:
We would like to thank the MDM TAC for generous time allocations,
without which
this work would not have been possible; Dave
Mills for his time and help with the Linux camera drivers; and
the anonymous referee for helpful comments and suggestions.
Support for DMT and MAH came in part from funds provided by the Ohio
State University Department of Astronomy.
This material is based in part upon work supported by the 
National Science Foundation under Grant Number AST007480.
Any opinions, findings, and conclusions or recommendations 
expressed in this material are those of the author(s) and do not necessarily
 reflect the views of the National Science Foundation.
SLH and JRE were partially funded by the Missouri Space Grant Consortium.


\begin{thebibliography}{}
\bibitem{breg1} Breger M., et al. 1994, A\&A, 289, 162
\bibitem[Charpinet et al. 2001]{char96}Charpinet, S., 
 Fontaine, G., \& Brassard, P. 2001, PASP, 113, 775
\bibitem{char02}Charpinet S., Fontaine G., Brassard P.,  Dorman, B.
2002, ApJS, 140, 469
\bibitem{charp05}Charpinet S., Fontaine G., Brassard P., Green E.M.
Chayer P. 2005, A\&A, 437, 575
\bibitem[JCD (2000)]{jcd00}Christensen-Dalsgaard J. 2004, SoPh,
220, 137
\bibitem[Green et al. 2003]{grn03}Green E.M., et al. 2003, ApJ, 583, L31
\bibitem{heb2}Heber U., Hunger K., Jonas G., Kudritzki R. P. 
1984, A\&A, 130, 119
\bibitem{kaw05}Kawaler S.D., Hostler S.R. 2005, ApJ, 621, 432
\bibitem{kaw1}Kawaler S.D., Vu\u{c}kovi\'{c}, M., The WET Collaboration 
2006, BaltA, 15, 283
\bibitem[Kilkenny et al 2001]{kil97}Kilkenny D. 2001, ASP Conf. Ser., 259,
356, IAU Colloquium 185, Radial and Nonradial Pulsations as Probes of 
Stellar Evolution, ed.
 C. Aerts, T. Bedding, \& J. Christensen-Dalsgaard (San Francisco:
 ASP), 356.
\bibitem{koen05}Koen C., O'Donoghue D., Kilkenny D.,  Pollacco D. L.
 2004, NewA, 9, 565
\bibitem{ost}\O stensen R., Heber U., Silvotti R., Solheim J.-E., 
Dreizler S.,   Edelmann H.  2001, A\&A, 378, 466O
\bibitem{simon}O'Toole S. J., Heber U., Benjamin R. A. 2004, A\&A,
422, 1053O
\bibitem{per}Pereira T.M.D., Lopes I.P. 2005, ApJ, 622, 1068
\bibitem[Pesnell (1985)]{pes85} Pesnell W. Dean, 1985, ApJ, 292, 238
\bibitem[Reed et al. (2005)]{me1}Reed M.D., Brondel B.J.,  Kawaler S.D.
2005, ApJ, 634, 602
\bibitem[Reed et al. (2003)]{me2}Reed M.D., et al. (The Whole Earth Telescope
Collaboration) 2004, MNRAS, 348, 1164
\bibitem[Saffer et al. 1994]{saf94}Saffer R.A., Bergeron P., Koester D., 
 Liebert J. 1994, ApJ, 432, 351
\bibitem[Winget et al. (1991)]{wing91}Winget D.E., et al. (The Whole Earth
Telescope Collaboration) 1991, ApJ, 378, 326
\bibitem{zhou}Zhou A.-Y., et al. 2006, MNRAS, 367, 179
\end{thebibliography}
\end{document}